\def\year{2022}\relax
\documentclass[letterpaper]{article} 
\usepackage{aaai22}  
\usepackage{times}  
\usepackage{helvet}  
\usepackage{courier}  
\usepackage[hyphens]{url}  
\usepackage{graphicx} 
\usepackage{natbib}  
\usepackage{subcaption}
\usepackage{caption} 
\DeclareCaptionStyle{ruled}{labelfont=normalfont,labelsep=colon,strut=off} 
\usepackage{xcolor}
\usepackage{algorithm}
\usepackage{algorithmicx}
\usepackage{booktabs}
\usepackage{amsmath,amsthm,amsfonts,amssymb}
\usepackage{bm}
\usepackage{multirow}
\usepackage{newfloat}
\usepackage{listings}
\floatstyle{ruled}
\newfloat{listing}{tb}{lst}{}
\floatname{listing}{Listing}
\newcommand\algcomment[1]{\def\@algcomment{\footnotesize#1}}
\def\onedot{.}
\def\eg{{\em e.g}\onedot} 
\def\ie{{\em i.e}\onedot}

\usepackage{float}

\def\modelname{{DiffSED}}
\def\modelnamebb{{DiffSED-BB}}
\newcommand{\red}[1]{\textcolor{black}{#1}}

\title{DiffSED: Sound Event Detection with Denoising Diffusion}
\author{
Swapnil Bhosale\textsuperscript{\rm 1*}, Sauradip Nag\textsuperscript{\rm 1*}, Diptesh Kanojia\textsuperscript{\rm 1}, Jiankang Deng\textsuperscript{\rm 2}, Xiatian Zhu\textsuperscript{\rm 1}
}
\affiliations{
    \textsuperscript{\rm 1}University of Surrey, UK\\
    \textsuperscript{\rm 2}Imperial College London, UK\\
    s.bhosale@surrey.ac.uk, s.nag@surrey.ac.uk
}
\usepackage{bibentry}

\begin{document}
\maketitle
\begin{abstract}

Sound Event Detection (SED) aims to predict the temporal boundaries of all the events of interest and their class labels,
given an unconstrained audio sample.
Taking either the split-and-classify (\ie, frame-level) strategy or 
the more principled event-level modeling approach, 
all existing methods consider the SED problem
from the discriminative learning perspective.
In this work, we reformulate the SED problem by taking a generative learning perspective.
Specifically, we aim to generate sound temporal boundaries from noisy proposals in a denoising diffusion process, conditioned on a target audio sample.
During training, our model learns to reverse the noising process by converting noisy latent queries to the ground-truth versions
in the elegant Transformer decoder framework.
Doing so enables the model generate accurate event boundaries from even noisy queries during inference.
%
Extensive experiments on the Urban-SED and EPIC-Sounds datasets demonstrate that our model significantly outperforms existing alternatives,
with 40+\% faster convergence in training.
\end{abstract}

\section{Introduction}
\label{sec:intro}
Sound event detection (SED) aims to temporally localize sound events of interest (\ie, the start and end time) and recognize their class labels in a long audio stream \cite{mesaros2021sound}. 
As a fundamental audio signal processing task, it has become the cornerstone of many related recognition scenarios, such as audio captioning \cite{xu2021text,bhosale2023novel,xie2023enhance}, and acoustic scene understanding \cite{igarashi2022information,bear2019towards}.

 \begin{figure}[t]
   \centering
   \includegraphics[width=1.05\linewidth]{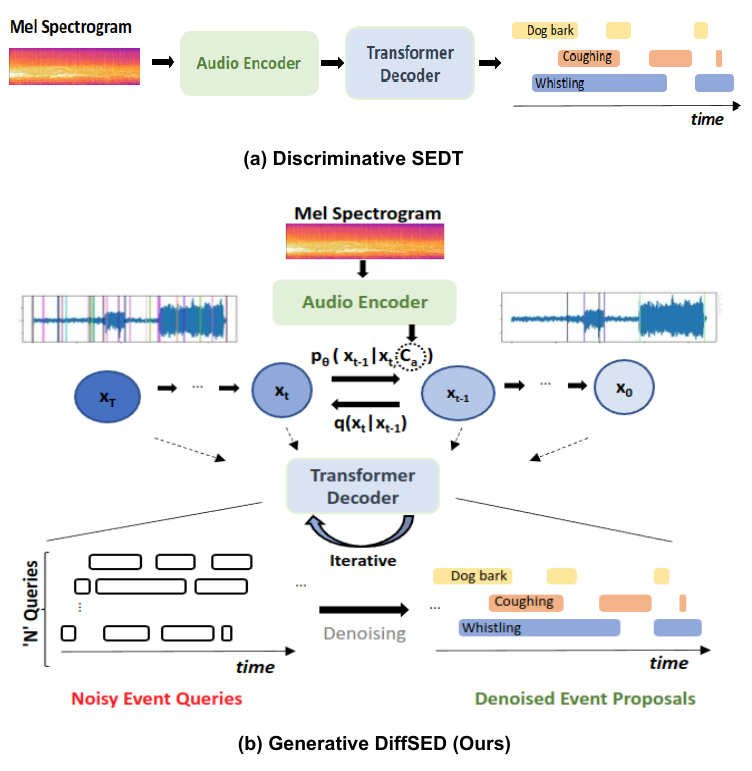}
   \caption{Architectural comparison:
   (a) Conventional discriminative DETR-based Sound Event Detector Transformer (SEDT) \cite{ye2021sound} 
 incorporates a single decoding step with clean queries. 
   (b) Our diffusion-infused generative DETR-based Sound Event Detector (\modelname) conducts multi-step decoding/denoising over noised queries.
   }
   \label{fig:fig1}
   \vspace{-1.5em}
 \end{figure}
In the literature, all existing SED methods can be grouped into two categories namely, frame-level and event-level approaches. 
Frame-level approaches classify each audio frame/segment into event classes and then aggregate the consecutive frame-level predictions to identify sound event boundaries or endpoints \cite{miyazaki2020convolution, lin2019guided}.
They are often heavily manually designed with plenty of heuristics and data-specific parameter optimization,
hence less scalable and reliable across different audio data.
Event-level approaches, on the other hand, directly model the temporal boundaries of sound events, taking into account the correlation between frames, thereby eliminating the mundane post-processing step and are more generalizable \cite{ye2021sound}.
In both approaches, existing methods 
rely on {\em proposal prediction} by regressing the start and end times of each, \ie, discriminative learning based.

Recently, generative learning models such as diffusion models \cite{ho2020denoising,song2020denoising} have emerged strongly in computer vision.
Conceptually, we draw an analogy between the SED problem and image-based object detection \cite{duan2019centernet,mmdetection}.
We consider the latest generative learning based object detection approach \cite{chen2022diffusiondet} represents a new direction for designing detection models in general. 
Although conceptually similar to object detection, the SED problem still presents unique challenges
and complexity due to the presence of temporal dynamics.
%
Besides, there are several limitations with the detection diffusion formulation in \cite{chen2022diffusiondet}. 
First, a two-stage pipeline (\eg, RCNN \cite{chao2018rethinking}) is adopted, giving rise to localization-error propagation from proposal generation to proposal classification \cite{nag2022gsm,nag2022pftm,nag2022pclfm}.
Second, as each event proposal is processed individually, their intrinsic relationship modeling is overlooked, potentially hurting the learning efficacy.
To address these issues, we present two different designs:
(a) Adopting the one-stage detection pipeline \cite{tian2019fcos,wang2020solo}
that have already shown excellent performance with a relatively simpler design, in particular, DETR \cite{carion2020end}.
Even within the SED literature, 
this simpler pipeline
has shown to achieve higher accuracy than frame-level models on a variety of sound event detection datasets due to the better temporal resolution, as well as its ability to learn long-range dependencies between sound events \cite{ye2021sound}.
(b) A unique challenge with SED is \textbf{\em big boundary ambiguity} as compared to object detection. 
This is because temporal audio events are continuous in time without clear start and end points (\eg, non-zero momentum), and the transition between consecutive events is often stochastic. Further, human perception of event boundaries is also instinctive and subjective. For the above reasons, we reckon that diffusion-based models could be a great fit for sound event detection.

Nonetheless, it is non-trivial to integrate denoising diffusion with existing sound event detection models, due to several reasons.
(1) 
Whilst efficient at processing high-dimension data simultaneously, diffusion models \cite{dhariwal2021diffusion,li2022diffusion} have typically been shown to work with continuous input data.
But event boundaries in SED are discrete.
(2) 
Denoising diffusion and SED
both suffer low efficiency, and their combination would even get worse.
Both of the problems have not been investigated systematically thus far.

To address the aforementioned challenges,
a novel {\em conditioned event diffusion} method is proposed 
for efficiently tackling the SED task, 
abbreviated as {\bf \modelname}.
In the forward diffusion process, 
Gaussian noises are added to the event latents iteratively.
In the reverse denoising process, the noisy latents
are passed as queries to a denoiser (\eg, DETR \cite{carion2020end}) for denoising the event latents so that desired event proposals can be obtained, with the condition on the observation of an input audio stream.
The usage of noisy latents allows our model
to bypass the need for continuous input,
as the denoising diffusion process takes place in the designated latent space.
During inference, the model can take as input the noisy latents
composed of noises sampled from Gaussian distribution and learned components, 
and outputs the event proposals of a given audio stream (\ie, the condition).
%
The proposed noise-to-queries strategy for denoising diffusion has several appealing properties:
(i) Evolutionary enhancement of queries during inference wherein each denoising step can be interpreted as a unique distribution of noise thus adding stochasticity to solve the boundary ambiguity problem. 
(ii) Integrating denoising diffusion with this noisy-latent decoder design solves the typical slow-convergence limitation.
We summarize the {\bf contributions} of this work.
{\bf(a)} We 
reformulate sound event detection (SED) as a generative denoising process in an elegant transformer decoder framework.
This is the first study to apply the diffusion model for the SED task to the best of our knowledge.
{\bf (b)} 
The proposed generative adaptation uses a noise-to-queries strategy with several appealing properties such as evolutionary enhancement of queries and faster convergence.
{\bf (c)} Our comprehensive experiments on the URBAN-SED \cite{salamon2014dataset} and the EPIC-Sounds \cite{EPICSOUNDS2023} datasets validate the significant performance advantage of our \modelname{} over existing alternatives.

\begin{figure*}[t]
    \centering
    \includegraphics[width=0.99\textwidth]{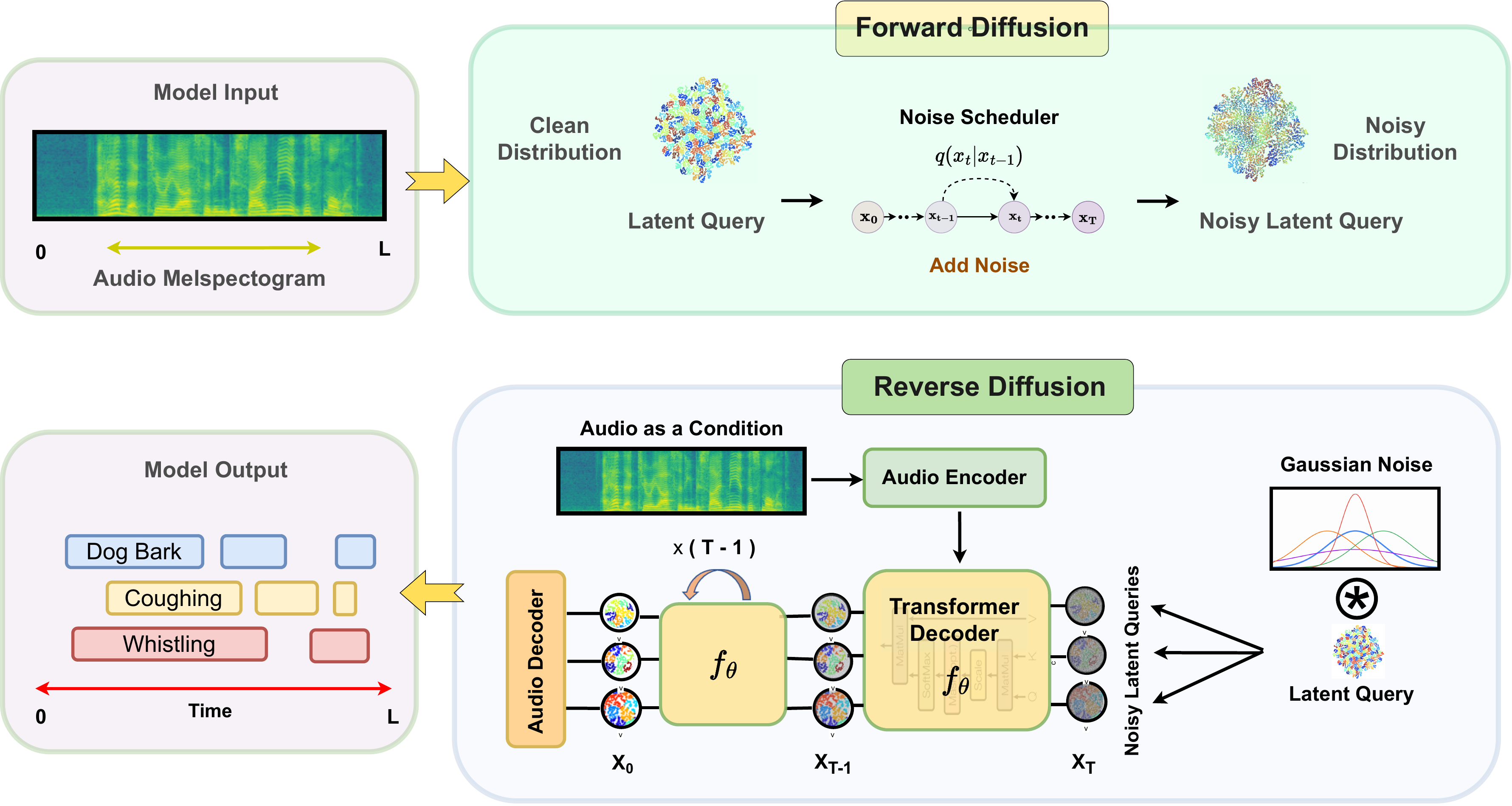}
    \caption{\textbf{Overview of our proposed \modelname}. ({\bf Top}) In the forward diffusion process, Gaussian noises are added to the event latents iteratively to obtain noisy latents $X_{T}$. 
    ({\bf Bottom}) In the reverse denoising process, an audio melspectogram is passed as the condition along with random noisy latents sampled from the Gaussian distribution. The noisy latents are passed as the query to the denoiser for denoising the event latents in an iterative fashion to obtain event proposals.}
    \label{fig:main}
\end{figure*}
\section{Related work}
\subsection{Sound event detection}
The existing SED literature can be divided into two categories, namely, frame-level approaches and event-level approaches. 
In frame-level approaches \cite{lim2017rare,turpault2019sound,miyazaki2020convolution}, the input audio signal is first divided into short, fixed-length segments, and the sound events within each segment are further classified {\em independently}. 
Despite strong performance and good intuition,
this split-and-classify strategy requires plenty of heuristics designs, unscalable parameter settings (\eg, segment duration),
as well as time-consuming post-processing (\eg, aggregating frame-level predictions).
To overcome these limitations,
event-level approaches \cite{ye2021sound} present a more principled and scalable solution
with end-to-end learning frameworks,
inspired by the model designs in object detection \cite{carion2020end,zhu2020deformable,zhang2022dino} and video action recognition domains \cite{tan2021relaxed,shi2022react}.
Whilst being understudied, this strategy has shown to be more efficient and robust to longer and more complex (overlapping) events, such as those in music and human speech as well as short and frequently occurring events such as those in urban soundscapes or environmental monitoring.
Our \modelname{} belongs to this category, further pushing this forefront of performance.


Deep learning techniques have achieved excellent performance in SED. For instance, convolutional neural networks (CNNs) have been widely investigated for audio event classification \cite{cakir2017convolutional,kumar2018knowledge} owing to their ability to efficiently capture and analyze local patterns within the acoustic waveform of sound. Additionally, recurrent neural networks (RNNs) have been used for temporal modeling of audio signals in arrears to their propensity to capture long-term temporal dependencies in sequential data - an innate property of audio signals. Interestingly, apart from the hybrid approaches \cite{li2020sound,koh2021sound}, that utilize CNNs to extract features from the audio signal, which are then fed into an RNN to model temporal dependencies, recently, transformer based architectures \cite{wakayama2022cnn,chen2022hts}
have been shown as equally promising, particularly, leveraging the self-attention mechanisms to model temporal relationships in audio signals and capturing complex patterns over time.
Commonly, all the prior methods consider the SED problem as discriminative learning.
In contrast, we treat for the first time this problem in a unique perspective of generative learning.
In particular, we generate the sound event bounds and predict the class labels from noise latents, with the condition to the input audio sample.
\vspace{-1em}
\subsection{Diffusion-based models for audio tasks}
As a new class of deep generative models, diffusion models have been gaining popularity in different fields. Beginning with a sample from a random distribution, the diffusion model is optimized to gradually learn a denoising schedule to obtain a noise-free target. This paradigm has yielded remarkable results in audio processing tasks ranging from audio generation \cite{leng2022binauralgrad,huang2022fastdiff}, audio enhancement \cite{lemercier2022analysing}, audio separation \cite{lutati2023separate} etc. 
To the best of our knowledge, this is the first work that exploits a diffusion model for the SED task.
    

    
\section{Methodology}

\noindent\textbf{Problem definition} Sound event detection (SED) involves both classification and {\em temporal} localization given an audio sequence. In this task, the audio sequence is usually represented as a 2-dimensional feature, such as a mel-spectrogram. We want a model to output the onset and offset times of all target events and the corresponding event labels \cite{wakayama2022cnn}. To train the model, we collect a set of labeled audio sequence set $D^{train} = \{A_{i}, \psi_{i}\}$. Each audio $A_{i} \in \mathcal{R}^{T \times F}$ (where $T \times F$ represents the spectro-temporal dimension) is labeled with temporal annotation $\psi_{i}= \{ (\Psi_{j}, \xi_{j}, y_{j})\}_{j=1}^{M_{i}}$ where $\Psi_{j} / \xi_{j}$ represents onset/offset of an event and $y_{j}$ denotes the acoustic class event label.


\subsection{Preliminaries on diffusion model}
Diffusion models are a class of generative models that use the diffusion process to model complex probability distributions \cite{ho2020denoising,song2020denoising}. In a diffusion model, the forward process generates samples by iteratively applying a diffusion equation to a starting noise vector. The forward process can be represented by the following equation:
\begin{equation}
\label{eq:forward}
z_t = \sqrt{1 - \beta_t} * z_{t-1} + \sqrt{\beta_t} * x_t
\end{equation}
where $z_t$ is the diffusion state at time $t$, $x_t$ is the input at time $t$, and $\beta_t$ is the diffusion coefficient at time $t$. 
The noise scale is controlled by $\beta_{t}$ which adopts a monotonically decreasing cosine schedule \cite{ho2020denoising,song2020denoising} in every different time step $t$. 

Denoising in diffusion models is the process of generating a clean representation from a noisy observation by reversing the diffusion process.
In other words, the goal is to obtain an estimate of the original representation from the final diffusion state. The denoising process can be performed using the reverse diffusion process, which can be represented by the following equation:
\begin{equation}
z_{T-t} = (z_{T-t+1} - \sqrt{\beta_{T-t}} * x_{T-t}) / \sqrt{1 - \beta_{T-t}}
\end{equation}
where $z_T$ is the final diffusion state, $x_t$ is the noisy input at time $t$, and $\beta_{T-t}$ is the diffusion coefficient at time $T-t$. The denoising process starts from the final diffusion state $z_T$ and iteratively applies the reverse diffusion equation to obtain an estimate of the original representation $x_0 = z_1$.
\begin{align}
 x_t = (z_{t+1} - \sqrt{\beta_t} * x_{t-1}) / \sqrt{1 - \beta_t}
\end{align}
where $\forall t \in [1, T-1]$ such that $x_t$ is the estimate of the original representation at time $t$. The denoising process can be improved by adding regularization or constraints to the estimate of the original representation.

\subsection{\modelname{}: Architecture design}
\textbf{Diffusion-based SED formulation} In this work, we formulate the SED task in a conditional denoising diffusion framework. In our setting, data samples are a set of learnable event query embeddings $\textbf{z}_{0} = b$, where $b \in \mathbb{R}^{N \times D}$ denotes $N$ event query embeddings at the dimension of $D$. In our implementation, the event queries are retrieved from a simple lookup table that stores embeddings of a fixed dictionary of size N (initialized from $\mathcal{N}(0, 1)$). A neural network $f_{\theta}(\textbf{z}_{t},t,A)$ is trained to predict $\textbf{z}_{0}$ from noisy proposals $\textbf{z}_{t}$, conditioned on the corresponding audio $A$. The audio category $\hat{y}$ is predicted subsequently. See Algorithm \ref{alg:train} for more details.

\begin{algorithm}[t]
	\caption{\small Training}
	\label{alg:train}
	\algcomment{\fontsize{7.2pt}{0em}\selectfont \textt{alpha\_cumprod(t)}: cumulative product of $\alpha_i$, \ie, $\prod_{i=1}^t \alpha_i$}

	\definecolor{codeblue}{rgb}{0.25,0.5,0.5}
	\definecolor{codegreen}{rgb}{0,0.6,0}
	\definecolor{codekw}{RGB}{207,33,46}
	\lstset{
		backgroundcolor=\color{white},
		basicstyle=\fontsize{7.5pt}{7.5pt}\ttfamily\selectfont,
		columns=fullflexible,
		breaklines=true,
		captionpos=b,
		commentstyle=\fontsize{7.5pt}{7.5pt}\color{codegreen},
		keywordstyle=\fontsize{7.5pt}{7.5pt}\color{codekw},
		escapechar={|}, 
	}
	\begin{lstlisting}[language=python]
	def train_loss(audio, event_query):
	"""
	audio: [B, T, F]
	event_queries: [B, N, D]
	# B: batch_size
	# N: number of event queries
	"""
	# Encode audio features
	audio_feats = audio_encoder(audio)
	
	# Signal scaling
	event_queries = (event_queries * 2 - 1) * scale  
	
	# Corrupt event_queries
	t = randint(0, T)|~~~~~~~~~~~|# time step
	eps = normal(mean=0, std=1)  # noise: [B, N, D]
	event_queries_crpt = sqrt(|~~~~|alpha_cumprod(t)) * event_queries + 
	|~|sqrt(1 - alpha_cumprod(t)) * eps
	
	# Predict bounding boxes
	pb_pred = detection_decoder(event_queries_crpt, audio_feats, t)
	
	# Set prediction loss
	loss = set_prediction_loss(pb_pred, gt_boxes)
	
	return loss
	\end{lstlisting}
\end{algorithm}
Since the diffusion model generates a data sample iteratively, it needs to run the model $f_{\theta}$ multiple times in inference. It would be computationally intractable to directly apply $f_{\theta}$ on the raw audio at every iterative step. For efficiency, we propose to separate the whole model into two parts, \textit{audio encoder} and \textit{detection decoder}, where the former runs only once to extract a feature representation of
the input audio $A_{i}$, and the latter takes this feature as a condition to progressively refine the noisy proposals $\textbf{z}_{t}$.

\noindent\textbf{Audio encoder}
The audio encoder takes as input the pre-extracted audio mel-spectograms and extracts high-level features for the following detection decoder. In general, any audio encoder can be used. We follow \cite{ye2021sound} for the audio encoder.
More specifically, the raw audio is first encoded using a CNN based encoder backbone (\ie, ResNet-50) to obtain the audio feature 
$A_{f} \in \mathbb{R}^{T^{'} \times F^{'}}$ respectively. 
This is followed by a multi-layered temporal transformer \cite{vaswani2017attention} $\tau$ that performs global attention across the time dimension to obtain the global feature as:
\begin{equation}
    C_{a} = \tau(A_{f})
\end{equation}
where query, key, and value of the transformer is set to $A_{f}$. We also append positional encoding to $A_{f}$ before passing it into the transformer.

\subsubsection{Detection decoder}
Similar to SEDT \cite{ye2021sound}, we use a transformer decoder \cite{vaswani2017attention} (denoted by $f_{\theta}$) for detection. Functionally, in our formulation it serves as a denoiser. In traditional DETR \cite{lin2021detr},  the queries are learnable continuous embeddings with random initialization. In \modelname{}, however, we exploit the queries
as the \textit{denoising targets}.

As opposed to adding noises to object boundaries \cite{chen2022diffusiondet,nag2023difftad}, we inject the Gaussian noise to the randomly initialized latent queries.
This is similar to the concept of \textit{event queries}  \cite{rombach2022high}. To detect multiple events occurring simultaneously, we sample $N$ such \textit{noisy event queries} to form $Q \in \mathbb{R}^{N \times D}$ which will be subsequently passed on to the detection decoder for denoising. 
Taking $Q$ as input, the decoder predicts $N$ outputs:
\begin{equation}
    F_{d} = f_{\theta}(Q;C_{a}) \in \mathbb{R}^{N \times D}
\end{equation}
where $C_{a}$ is the encoded audio feature and the $F_{d}$ is the final embedding. $F_{d}$ is finally decoded using two parallel heads
namely (1) event classification head and (2) event localization head respectively. The
first estimates the probability of a particular event within
the event proposal.  The second estimates the onset and offset of event in the raw audio.

\subsection{Model training}
During training, we first construct the diffusion process
that corrupts the event latents to noisy latents. We then train the model to reverse this noising process.
We add Gaussian noises to the learnable queries.
The noise scale is controlled by $\beta_{t}$ (Eq. \eqref{eq:forward}), which adopts a monotonically decreasing cosine schedule in different timestep $t$, following \cite{ho2020denoising, song2020denoising}.
The decoder uses the noisy event queries (corresponding to $t$) and the global feature $C_a$ as the condition (see Fig \ref{fig:fig1} (b)) to generate the denoised event queries (corresponding to $t-1$) repeatedly until an approximation of $Q$ is obtained. The output from the last denoising step (corresponding to each input event query) is projected into sigmoidal onset and offset timestamps and an event probability distribution using separate feedforward projection layers.
We observe that SED favors a relatively high signal scaling
value than object detection \cite{chen2022diffusiondet} (see Table \ref{tab:ablation_noise_scale}).

The event-based objective is defined as a combination of a binary classification loss for event onset and offset prediction and a cross-entropy loss for event class prediction.
We compute Hungarian assignment between ground truth boxes and the outputs of the model. We supervise the model training using each pair of matched ground-truth/prediction (event class and the temporal boundary).

\subsection{Model Inference}
In inference, the noisy event queries are randomly sampled from a Gaussian distribution.
Starting from noisy latents sampled from a Gaussian distribution, the model progressively refines the predictions. 
At each sampling step, the random or estimated latents from the last sampling step are sent into the detection decoder to predict the event category and the event onset/offsets. After obtaining the event proposals of the current step, DDIM~\cite{song2021denoising} is adopted to estimate the proposals for the next step.
DiffSED has a simple event proposal
generation pipeline without post-processing (\eg, non-maximum suppression). 

\subsection{Key Insights}
\paragraph{One model multiple trade-offs}
Once trained,
\modelname{} works under a varying number of event queries and sampling steps in inference.
While inferring, each sampling step involves, estimating event queries from the last sampling step and sending them back into the detection decoder to eventually predict the event classes and event boundaries at the $t_0$ step, \ie, fully denoised.
In general, better accuracy can be obtained using more queries and fewer steps (see Table \ref{tab:ablation_queries} and Table \ref{tab:multistep_decoding}). We discuss the multistep decoding experiments in detail in our ablation study.
Ultimately, it can be determined that a single \modelname{} can meet a number of different trade-off needs between speed and accuracy.
 \begin{figure}[h]
   \centering
   \includegraphics[width=1.05\linewidth]{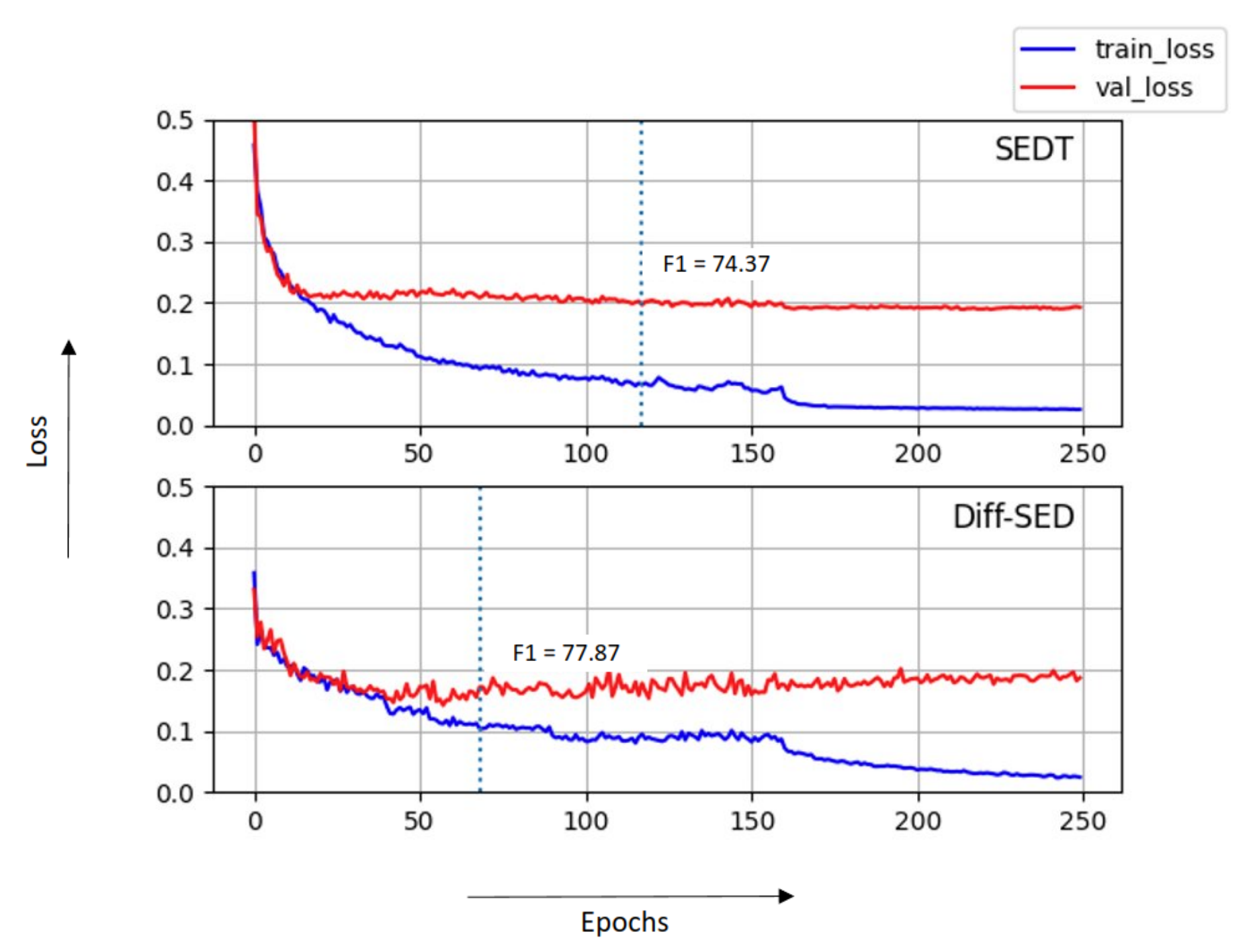}
   \caption{\red{Convergence rates for SEDT and \modelname{} on the URBAN-SED dataset. The dotted lines represent the training epoch when the best-performing checkpoint (the one with the best audio-tagging F1 score on the validation set) arrived. 
   \modelname{} trains faster ($>$40\%) and achieves better optimum than SEDT.
   }}
   \label{fig:fig3}
 \end{figure}

\paragraph{Faster convergence}
DETR-style detection models suffer generally slow convergence \cite{liu2022dabdetr} due to inconsistent matching of event queries to the event proposals. Concretely, for the same audio, an event query is often matched with different event boundaries in different epochs, making the optimization oscillating and difficult.
%
In \modelname{} each query is designed as a proposal proxy -- a noised event query that can be regarded as a good event proposal due to staying close to the corresponding ground truth boundary. 
Our query denoising task thus has a definite optimization objective which is the ground truth proposal.
%
We validate that query denoising based \modelname{} converges faster than SEDT (see Fig \ref{fig:fig3}),
whilst achieving superior performance (Table \ref{tab:urbansed}).

\begin{algorithm}[t]
	\caption{\small Noise corruption}
	\label{alg:denoising}	
	\definecolor{codeblue}{rgb}{0.25,0.5,0.5}
	\definecolor{codegreen}{rgb}{0,0.6,0}
	\definecolor{codekw}{RGB}{207,33,46}
	\lstset{
		backgroundcolor=\color{white},
		basicstyle=\fontsize{7.5pt}{7.5pt}\ttfamily\selectfont,
		columns=fullflexible,
		breaklines=true,
		captionpos=b,
		commentstyle=\fontsize{7.5pt}{7.5pt}\color{codegreen},
		keywordstyle=\fontsize{7.5pt}{7.5pt}\color{codekw},
		escapechar={|}, 
	}
	\begin{lstlisting}[language=python]
	def add_noise():
	"""
	gt_boxes: [B, *, 2]
	event_queries: [B, N, D]
	B: batch_size
	N: number of event queries
	"""
	if corrupt bounding_boxes: # Diff-SED-BB
        # Padding (repeat) bounding boxes
        pb = Pad(gt_boxes, N) #[B, N, 2]
        # Signal scaling
    	pb = (pb * 2 - 1) * scale  	
    	# Corrupt bounding boxes
    	t = randint(0, T)|~~~~~~~~~~~|#time step
    	eps = normal(mean=0, std=1)  #noise: [B, N, 2]
    	pb_crpt = sqrt(alpha_cumprod(t)) * pb + sqrt(1 - alpha_cumprod(t)) * eps
        event_queries_crpt = Project(pb_crpt) 
        #[B, N, 2] -> [B, N, D]
    else: # DiffSED
	    # Signal scaling
    	event_queries = (event_queries * 2 - 1) * scale  	
    	# Corrupt event_queries
    	t = randint(0, T)|~~~~~~~~~~~|#time step
    	eps = normal(mean=0, std=1)  #noise: [B, N, D]
    	event_queries_crpt = sqrt(alpha_cumprod(t)) * event_queries + 
    	|~|sqrt(1 - alpha_cumprod(t)) * eps
	return event_queries_crpt
	\end{lstlisting}
\end{algorithm}

\section{Experiments}
\paragraph{Datasets}

We present our results on two datasets namely, URBAN-SED \cite{salamon2014dataset} and EPIC-Sounds \cite{EPICSOUNDS2023}. {\em URBAN-SED} is a publicly available dataset for SED in urban environments. It is accompanied by detailed annotations, including onset and offset times for each sound event, along with human generated accurate annotations.
The {\em EPIC-Sounds} dataset consists of more than 36,000 audio recordings of various lengths, totaling over 500 hours of audio. The recordings were made in a variety of indoor and outdoor environments, including office spaces, public places, and natural environments. They cover a wide range of sound classes, including human speech, animal sounds, environmental sounds, and music.

\begin{table*}[t]
    \centering
    \caption{Results on URBAN-SED (Test set)}
    \begin{tabular}{c|ccc|ccc|c}
    
    \toprule
    Model & \multicolumn{3}{c}{Event-based $[\%]$}&\multicolumn{3}{c}{Segment-based$[\%]$}&Audio tagging$[\%]$\\
     &$\rm{F1}$&$\rm{P}$&$\rm{R}$&$\rm{F1}$&$\rm{P}$&$\rm{R}$&$\rm{F1}$ \\
    \midrule\midrule
    CRNN-CWin \cite{miyazaki2020weakly} & $36.75$ & $-$ & $-$ & $65.74$ & $-$ & $-$ & $74.19$ \\
    Ctrans-CWin \cite{miyazaki2020weakly} & $34.36$ & $-$ & $-$ & $64.73$ & $-$ & $-$ & $74.05$ \\
     SEDT \cite{ye2021sound} & $37.27$ & $43.32$ & $33.21$ & $65.21$ & $74.82$ & $58.46$ & $74.37$ \\
     \midrule
     \modelname{} ({\bf Ours}) & \bf{43.89} & \bf{48.46} & \bf{37.82} & \bf{69.24} &	\bf{77.49} & \bf{62.05} &	\bf{77.87}\\
    \bottomrule
    \end{tabular}
    
    \label{tab:urbansed}
\end{table*}

\begin{table*}[t]
    \centering
     \caption{Results on EPIC-Sounds (Validation set)}
    \begin{tabular}{c|ccccc}
    \toprule
    
    Model & \rm{Top-1} & \rm{Top-5} & \rm{mCA} & \rm{mAP} & \rm{mAUC} \\ \midrule\midrule
    ASF {\cite{kazakos2021slow}} & 53.47 & 84.56 & 20.22 & 0.235 & \bf{0.879} \\
    SSAST {\cite{gong2022ssast}} & 53.75 & 84.54 & 20.11 & 0.254 & 0.873 \\\midrule
    \modelname{} ({\bf Ours}) & \bf{56.85} & \bf{87.45} & \bf{20.75} & \bf{0.277} & 0.861 \\
     
    \bottomrule
    \end{tabular}
    \label{tab:epicsounds}
\end{table*}

\paragraph{Evaluation metrics} 
To evaluate the model's performance on the URBAN-SED dataset, we measure F1-score, precision, and recall for both \texttt{event-level} and \texttt{segment-level} settings on the \texttt{test} split. For the EPIC-Sounds dataset, we report the top-1 and top-5 accuracy, as well as mean average precision (mAP), mean area under ROC curve (mAUC), and mean per class accuracy (mCA) on the \texttt{validation} split, following the protocol of \cite{EPICSOUNDS2023}.

\subsection{Implementation Details}
\paragraph{Training schedule}
We use a pre-trained encoder backbone ResNet-50 for feature extraction, for fair comparisons with previous methods \cite{ye2021sound}. Our model is trained for 400 epochs, while re-initializing the weights from the best checkpoint for every 100 epochs, using Adam optimizer with an initial learning rate of $10^{-4}$ with a decay schedule of $10^{-2}$. The batch size is set to 64 for URBAN-SED and 128 for EPIC-Sounds. All models are trained with 2 NVIDIA-A5500 GPUs. 
\paragraph{Testing schedule}
At the inference stage, the detection decoder iteratively refines the predictions from Gaussian random latent queries. 
For efficiency, by default, we denoise for a single time-step, \ie, $T_0 \leftarrow T_{1000}$ timestep. 


\subsection{Main Results}

\paragraph{Results on URBAN-SED}
We compare our model with previous end-to-end approaches under the supervised learning setting. The primary contribution of our work lies in proposing a diffusion-infused transformer decoder that provides a more robust representation of grounded event boundaries in the encoded acoustic features. From Table \ref{tab:urbansed}, we draw the following conclusions:
(1) The diffusion-based decoder of {\modelname{}} performs significantly better than all the other methods for both event-level and segment-level metrics, with a 6.62\% and 4.03\% absolute improvement, respectively. (2) Additionally, our model outperforms existing approaches in terms of audio-tagging results, with a 3.5\% absolute improvement. This validates our model formulation in exploiting the SED problem as generative learning in the denoising diffusion framework.

\paragraph{Results on EPIC-Sounds}
We use the publicly available pre-trained backbones ASF \cite{kazakos2021slow} and SSAST \cite{gong2022ssast} as competing models. We observe from Table \ref{tab:epicsounds} that:
(1) \modelname{} consistently outperforms both the alternatives with 3.1\% and 2.89\% improvement in the Top-1 and Top-5 accuracies, respectively; (2) Our model performs competitively in the mAUC score.

\subsection{Ablation study}
We conduct ablation experiments on URBAN-SED to study \modelname{} in detail. All experiments use the pre-trained ResNet-50 backbone features for training and inference without further specification.


\paragraph{Denoising strategy} Due to the inherent query
based design with the detection decoder, we discuss and compare two denoising strategies:
(1) Corrupting the event latents in the continuous space and passing it as queries 
(referred as \modelname{}, our choice).
(2) Corrupting discrete event proposals (\ie, ground-truth bounding boxes) 
and projecting it as queries 
(denoted as \texttt{\modelnamebb{}}, detailed in Algorithm \ref{alg:denoising}).
Additionally, we corrupt the label queries using random shuffle as the noise in the
forward diffusion step. 
To evaluate the effect of the denoising strategy experimentally, we
test both variants using different numbers of event proposals.
It can be observed in Table \ref{tab:ablation_queries} that both variants achieve the best audio-tagging performance when using 30 event proposals as input to the decoder. Also, the overall scores in both event-level and segment-level metrics are lesser for \modelnamebb{} compared to \modelname{}. We hypothesize this is caused by some adversarial effect 
in projecting the ground-truth bounding box (2-dimensional) to the latent event query.

\begin{table}[h]
    \centering
    \small
    \caption{Effect of the number of queries on the performance for URBAN-SED Test set. (AT: Audio Tagging performance)}
    \begin{tabular}{c|c|c|c|c}
    \toprule
    & \#Queries & \multicolumn{1}{c}{Event-F1$[\%]$}&\multicolumn{1}{c}{Segment-F1$[\%]$}&AT$[\%]$\\
    \midrule\midrule
    
    \parbox[t]{2mm}{\multirow{5}{*}{\rotatebox[origin=c]{90}{\modelnamebb{}}}}&10 & $31.43$ 
    & $58.85$ 
    & $68.87$ \\
    &20 & $35.32$ 
    & $60.53$ 
    & $68.84$ \\
     &30 & \textbf{37.29} 
     & \textbf{60.91} 
     & \textbf{69.61} \\
     &40 & $31.95$ 
     & $58.79$ 
     & $68.41$\\
     &50 & $31.81$ 
     & $57.89$ 
     & $68.31$\\
    \bottomrule
    
    \parbox[t]{2mm}{\multirow{5}{*}{\rotatebox[origin=c]{90}{\modelname{}}}}&10 & $40.78$ 
    & $68.41$ 
    & $77.22$ \\
    &20 & \textbf{41.42} 
    & \textbf{68.73} 
    & $76.54$ \\
     &30 & $41.3$ 
     & $68.21$ 
     & \textbf{77.46} \\
     &40 & $38.65$ 
     & $67.21$ 
     & $75.21$\\
     &50 & $36.28$ 
     & $64.22$ 
     & $72.77$\\
    \bottomrule
    \end{tabular}
    
    \label{tab:ablation_queries}
\end{table}

\paragraph{Multistep decoding}
\label{ablation:multistep}
We tabulate the results upon varying the number of denoising steps for both \modelname{} and \modelnamebb{} in Table \ref{tab:multistep_decoding}. We observe a steady improvement over the event-level and segment-level F1 scores as we increase the number of denoising steps from 1 to 5 and then gradually decrease when using 10 decoding steps. However, the best audio tagging performance is achieved when performing a single-step decoding. We hypothesize this is primarily because the event boundaries have short-range temporal dependencies that might not benefit significantly from multistep denoising. The noise addition mainly affects each time step independently and doesn't accumulate over multiple steps hence does not yield substantial improvements. Denoising over multiple timesteps requires more computing, 
while providing only a marginal gain 
thus not worthwhile.
\begin{table}[h]
    \centering
    \small
    \caption{Effect of the number of denoising steps used while inference on the performance for URBAN-SED Test set. (AT: Audio Tagging performance)}
    \begin{tabular}{c|c|c|c|c}
    \toprule
    & \#steps & \multicolumn{1}{c}{Event-F1$[\%]$}&\multicolumn{1}{c}{Segment-F1$[\%]$}&AT$[\%]$\\
    \midrule\midrule
    
    \parbox[t]{2mm}{\multirow{3}{*}{\rotatebox[origin=c]{90}{\tiny 
    \modelnamebb{}}}}&1 & $\textbf{39.78}$
    & $64.74$ 
    & $\textbf{72.92}$ \\
    &5 & $38.27$ 
    & $\textbf{65.72}$ 
    &  71.88\\
     &10 & $38.3$ 
     & $64.82$ 
     & $72.17$ \\
    \bottomrule
    
    \parbox[t]{2mm}{\multirow{3}{*}{\rotatebox[origin=c]{90}{\tiny \modelname{}}}}&1 & $43.89$ 
    & $69.24$ 
    & $\textbf{77.87}$ \\
    &5 & $\textbf{44.35}$
    & $\textbf{70.75}$ 
    & $77.07$ \\
     &10 & $43.50$ 
     & $69.05$ 
     & $77.36$ \\
    \bottomrule
    \end{tabular}
    
    \label{tab:multistep_decoding}
\end{table}

    


\paragraph{Signal scaling}
The signal scaling factor controls the signal-to-noise ratio~(SNR) of the diffusion process. 
We study the influence of scaling factors. 
The results in Table \ref{tab:ablation_noise_scale} demonstrate that the scaling factor of 0.4 achieves the highest audio-tagging performance as well as all other metrics for \modelname{}, whereas for \modelnamebb{} the best audio tagging performance is obtained for a scaling factor of 0.2 whilst achieving the best event-level and segment-level F1 score for a scaling factor of 0.4.
This suggests the relationship between optimal scaling and the denoising strategy.

\begin{table}[h]
    \centering
    \small
    \caption{Effect of scaling the noise factor on the performance for URBAN-SED Test set. (AT: Audio Tagging Performance)}
    \begin{tabular}{c|c|c|c|c}
    \toprule
    & Noise scale & \multicolumn{1}{c}{Event-F1$[\%]$}&\multicolumn{1}{c}{Segment-F1$[\%]$}&AT$[\%]$\\
    \midrule\midrule
    
    \parbox[t]{2mm}{\multirow{5}{*}{\rotatebox[origin=c]{90}{\modelnamebb{}}}}&0.1 & $32.61$
    & $32.45$ 
    & $73.49$ \\
    &0.2 & $35.91$ 
    & $35.73$ 
    & \textbf{75.73} \\
     &0.3 & $37.29$ 
     & $60.91$ 
     & $69.61$ \\
     &0.4 & \textbf{39.78} 
     & \textbf{64.74} 
     & $72.92$\\
     &0.5 & $33.14$ 
     & $61.79$ 
     & $71.12$\\
    \bottomrule
    
    \parbox[t]{2mm}{\multirow{5}{*}{\rotatebox[origin=c]{90}{\modelname{}}}}&0.1 & $37.61$ 
    & $54.63$ 
    & $72.2$ \\
    &0.2 & $39.65$ 
    & $58.17$ 
    & $73.89$ \\
     &0.3 & $41.3$ 
     & $68.21$ 
     & $77.46$ \\
     &0.4 & \textbf{43.89} 
     & \textbf{69.24} 
     & \textbf{77.87}\\
     &0.5 & $39.23$ 
     & $59.25$ 
     & $72.78$\\
    \bottomrule
    \end{tabular}
    
    \label{tab:ablation_noise_scale}
\end{table}

\begin{table}[H]
    \centering
    \small
    \caption{Effect of changing the seed value for inducing noise during inference. 
    Values inside $(.)$ indicate deviation from the mean calculated over 3 runs. }
    \begin{tabular}{c|c|c|c|c}
    \toprule
    & Runs & \multicolumn{1}{c}{Event-F1$[\%]$}&\multicolumn{1}{c}{Segment-F1$[\%]$}&AT$[\%]$\\
    \midrule\midrule
    
    \parbox[t]{1mm}{\multirow{4}{*}{\rotatebox[origin=c]{90}{\footnotesize \modelnamebb{}}}}&1 & $38.6 (\uparrow0.2)$ & $64.32 (\downarrow0.09)$ & $72.48 (0.0)$ \\
    &2 & $39.45 (\downarrow0.57)$ & $64.15 (\uparrow0.07)$ & $72.88 (\downarrow0.4)$ \\
     &3 & $38.57 (\uparrow0.3)$ & $64.21 (\uparrow0.01)$ & $72.08 (\uparrow0.4)$ \\
     \cmidrule{2-5}
     &\bf{Avg} & \bf{38.87} & \bf{64.22} & \bf{72.48} \\
     \midrule
    
    \parbox[t]{2mm}{\multirow{4}{*}{\rotatebox[origin=c]{90}{\modelname{}}}}&1 & $43.12 (\downarrow0.2)$ & $68.38 (\uparrow0.5)$ & $77.62 (\downarrow0.01)$ \\
    &2 & $42.35 (\uparrow0.5)$ & $68.97 (\uparrow0.01)$ & $77.59 (\uparrow0.02)$ \\
     &3 & $43.29 (\downarrow0.3)$ & $69.54 (\downarrow0.5)$ & $77.62 (\downarrow0.01)$ \\
     \cmidrule{2-5}
     &\bf{Avg} & \bf{42.92} & \bf{68.96} & \bf{77.61} \\
    \bottomrule
    \end{tabular}
    
    \label{tab:ablation_random_seeds}
\end{table}

\paragraph{Random seed}
\modelname{} starts with random noisy event queries
as input during inference. 
We evaluate the stability of \modelname{} and \modelnamebb{} by training three models independently with strictly the same configurations (30 noisy event proposals as input to the decoder and a scaling factor of 0.4) except for random seed on URBAN-SED dataset. Then, we evaluate each model instance with 3 different random seeds to measure the distribution of performance, inspired by \cite{chen2022diffusiondet,nag2023difftad}. As shown in Table \ref{tab:ablation_random_seeds}, most evaluation results are distributed closely to the average metrics for both variants.
This demonstrates that our models are robust to random event queries.

\section{Conclusion}
In this work, we reformulate the Sound Event Detection (SED) problem from the generative learning perspective, in particular under the
diffusion-based transformer framework.
We introduce a diffusion adaptation method
characterized by noisy event latents denoising.
This design has the advantage of being able to model the global dependencies of sound events, while still being computationally efficient. 
Our study verifies the efficacy of diffusion models in a new problem context (\ie, SED), consistent with previous findings.
%
Experiments show that our method is superior to existing art alternatives on standard benchmarks.
%

\bibliography{main}

\bigskip

\end{document}